\documentclass[sigplan,authorversion,nonacm]{acmart}

\usepackage{braket}
\usepackage[skip=1pt]{caption}
\usepackage{xcolor}
\usepackage{array}
\usepackage{listings}

\usepackage{caption}
\usepackage{subcaption}

\definecolor{darkgreen}{rgb}{0,0.5,0}
\definecolor{tinygray}{rgb}{0.94,0.94,0.94}
\usepackage{amsmath}

\makeatletter
\def\Decl@Mn@Delim#1#2#3#4{%
  \if\relax\noexpand#1%
    \let#1\undefined
  \fi
  \DeclareMathDelimiter{#1}{#2}{#3}{#4}{#3}{#4}}
\def\Decl@Mn@Open#1#2#3{\Decl@Mn@Delim{#1}{\mathopen}{#2}{#3}}
\def\Decl@Mn@Close#1#2#3{\Decl@Mn@Delim{#1}{\mathclose}{#2}{#3}}
\DeclareFontFamily{OMX}{MnSymbolE}{}
\DeclareFontShape{OMX}{MnSymbolE}{m}{n}{
    <-6>  MnSymbolE5
   <6-7>  MnSymbolE6
   <7-8>  MnSymbolE7
   <8-9>  MnSymbolE8
   <9-10> MnSymbolE9
  <10-12> MnSymbolE10
  <12->   MnSymbolE12}{}
\DeclareFontShape{OMX}{MnSymbolE}{b}{n}{
    <-6>  MnSymbolE-Bold5
   <6-7>  MnSymbolE-Bold6
   <7-8>  MnSymbolE-Bold7
   <8-9>  MnSymbolE-Bold8
   <9-10> MnSymbolE-Bold9
  <10-12> MnSymbolE-Bold10
  <12->   MnSymbolE-Bold12}{}
\DeclareSymbolFont{mnsymbols}  {OMX}{MnSymbolE}{m}{n}
\Decl@Mn@Open {\lsem}               {mnsymbols}{'102}
\Decl@Mn@Close{\rsem}               {mnsymbols}{'107}
\Decl@Mn@Open {\llangle}            {mnsymbols}{'164}
\Decl@Mn@Close{\rrangle}            {mnsymbols}{'171}
\makeatother

\usepackage{amsmath}

\AtBeginDocument{%
  \providecommand\BibTeX{{%
    \normalfont B\kern-0.5em{\scshape i\kern-0.25em b}\kern-0.8em\TeX}}}

\newcommand{\pnl}{
\affiliation{%
  \institution{Pacific Northwest National Laboratory}
  \city{Richland}
  \state{WA}
  \country{USA}
  \postcode{99354}
}}

\newcommand{\ornl}{
\affiliation{%
  \institution{Oak Ridge National Laboratory}
  \city{Oak Ridge}
  \state{TN}
  \country{USA}
  \postcode{37831}
}}

\newcommand{\microsoft}{
\affiliation{%
  \institution{Microsoft Research}
  \city{Redmond}
  \state{WA}
  \country{USA}
  \postcode{98052}
}}

\begin{document}

\title{TANQ-Sim: Tensorcore Accelerated Noisy Quantum  System Simulation via QIR on Perlmutter HPC}

\author{Ang Li}
\email{ang.li@pnnl.gov}
\orcid{0000-0003-3734-9137}
\pnl

\author{Chenxu Liu}
\email{chenxu.liu@pnnl.gov}
\orcid{0000-0003-2616-3126}
\pnl

\author{Samuel Stein}
\email{samuel.stein@pnnl.gov}
\orcid{0000-0002-2655-8251}
\pnl

\author{In-Saeng Suh}
\email{suhi@ornl.gov}
\orcid{0000-0002-6923-6455}
\ornl

\author{Muqing Zheng}
\email{muqing.zheng@pnnl.gov}
\orcid{0000-0002-6659-9672}
\pnl

\author{Meng Wang}
\email{meng.wang@pnnl.gov}
\orcid{0009-0008-1749-7929}
\pnl

\author{Yue Shi}
\email{yue.shi@pnnl.gov}
\orcid{0000-0002-1570-4444}
\pnl

\author{Bo Fang}
\email{bo.fang@pnnl.gov}
\orcid{0000-0001-9721-3982}
\pnl

\author{Martin Roetteler}
\email{martinro@microsoft.com}
\orcid{0000-0003-0234-2496}
\microsoft

\author{Travis Humble}
\email{humblets@ornl.gov}
\orcid{0000-0002-9449-0498}
\affiliation{%
  \institution{Quantum Science Center}
  \institution{Oak Ridge National Laboratory}
  \city{Oak Ridge}
  \state{TN}
  \country{USA}
  \postcode{37831}
}

\begin{abstract}
Although there have been remarkable advances in quantum computing (QC), it remains crucial to simulate quantum programs using classical large-scale parallel computing systems to validate quantum algorithms, comprehend the impact of noise, and develop resilient quantum applications. This is particularly important for bridging the gap between near-term noisy-intermediate-scale-quantum (NISQ) computing and future fault-tolerant quantum computing (FTQC). Nevertheless, current simulation methods either lack the capability to simulate noise, or simulate with excessive computational costs, or do not scale out effectively.

In this paper, we propose TANQ-Sim, a full-scale density matrix based simulator designed to simulate practical deep circuits with both coherent and non-coherent noise. To address the significant computational cost associated with such simulations, we propose a new density-matrix simulation approach that enables TANQ-Sim to leverage the latest double-precision tensorcores (DPTCs) in NVIDIA Ampere and Hopper GPUs. To the best of our knowledge, this is the first application of double-precision tensorcores for non-AI/ML workloads. To optimize performance, we also propose specific gate fusion techniques for density matrix simulation. For scaling, we rely on the advanced GPU-side communication library NVSHMEM and propose effective optimization methods for enhancing communication efficiency. Evaluations on the NERSC Perlmutter supercomputer demonstrate the functionality, performance, and scalability of the simulator. We also present three case studies to showcase the practical usage of TANQ-Sim, including teleportation, entanglement distillation, and Ising simulation. TANQ-Sim will be released on GitHub.

\end{abstract}

\maketitle
\pagestyle{plain} %

\section{Introduction}

Quantum computing (QC) \cite{nielsen2002quantum} is one of the most promising computing paradigms to address complex computing challenges that are classically intractable, such as those encountered in chemistry~\cite{georgescu2014quantum, kandala2017hardware}, cryptography~\cite{shor1999polynomial, gisin2002quantum}, machine learning~\cite{schuld2015introduction, biamonte2017quantum}, linear algebra~\cite{harrow2009quantum, clader2013preconditioned}, networking~\cite{kimble2008quantum, munro2010quantum}, and other fields. QC makes use of phenomena such as superposition and entanglement to perform computations, aiming at providing significant speedups over traditional high-performance computing resources, including the recently deployed exascale supercomputers.

Despite its potential, QC is still in its nascency, known as the \emph{Noisy Intermediate Scale Quantum} (NISQ) era~\cite{preskill2018quantum}, where only a limited number of physical qubits, typically ranging from a few dozen to a few hundreds, are available without \emph{Quantum Error Correction} (QEC)~\cite{lidar2013quantum}. The fundamental challenge with NISQ devices is that quantum devices are highly susceptible to noise, which can severely compromise their functionality and the accuracy of the computed results. Consequently, a key research challenge faced by the quantum computing community is to understand the impact of noise on contemporary NISQ devices, developing effective strategies to mitigate and tackle noise.

\begin{figure}[t!]
\minipage{0.49\columnwidth}
\includegraphics[width=0.9\columnwidth]{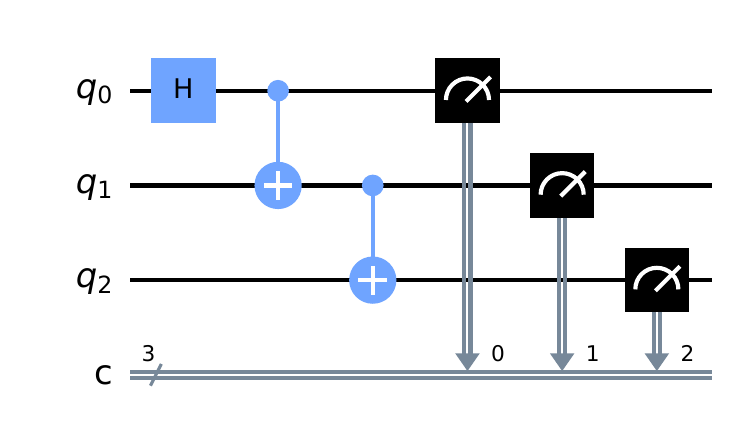} 
\endminipage\hfill
\minipage{0.49\columnwidth}
\includegraphics[width=0.9\columnwidth]{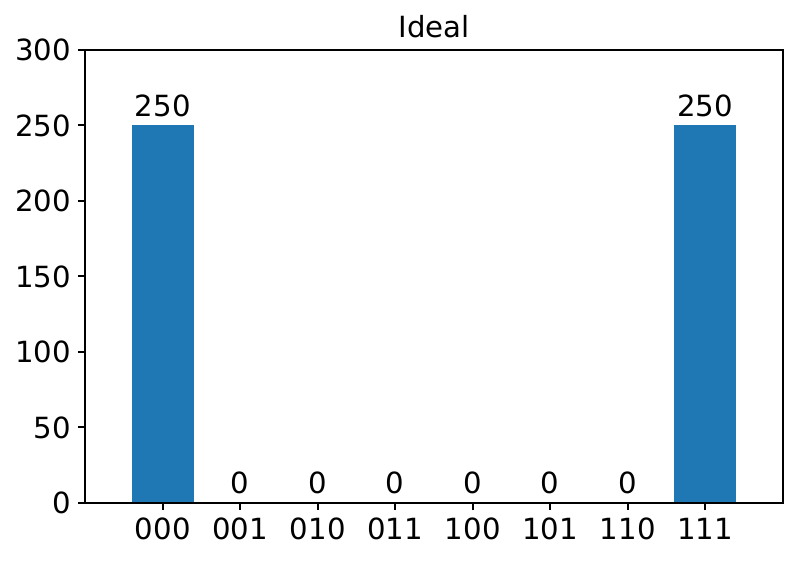} 
\endminipage

\minipage{0.49\columnwidth}
\includegraphics[width=0.9\columnwidth]{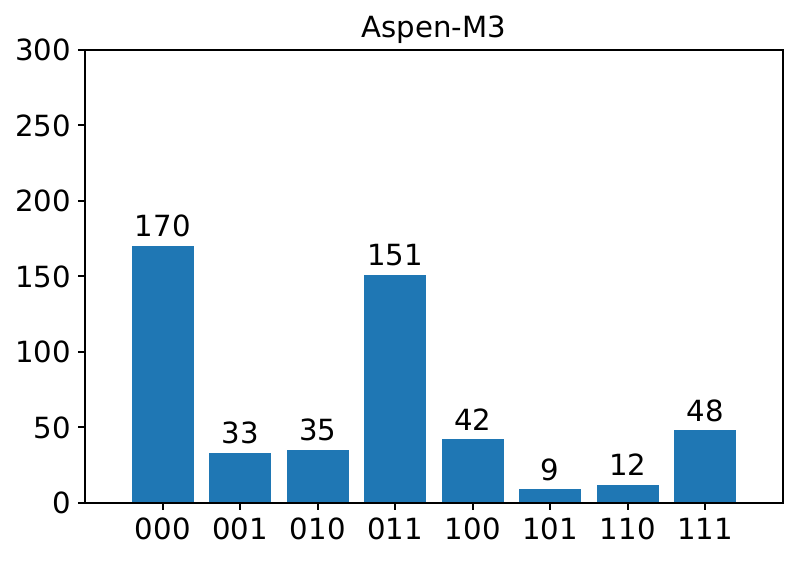} 
\endminipage\hfill
\minipage{0.49\columnwidth}
\includegraphics[width=0.9\columnwidth]{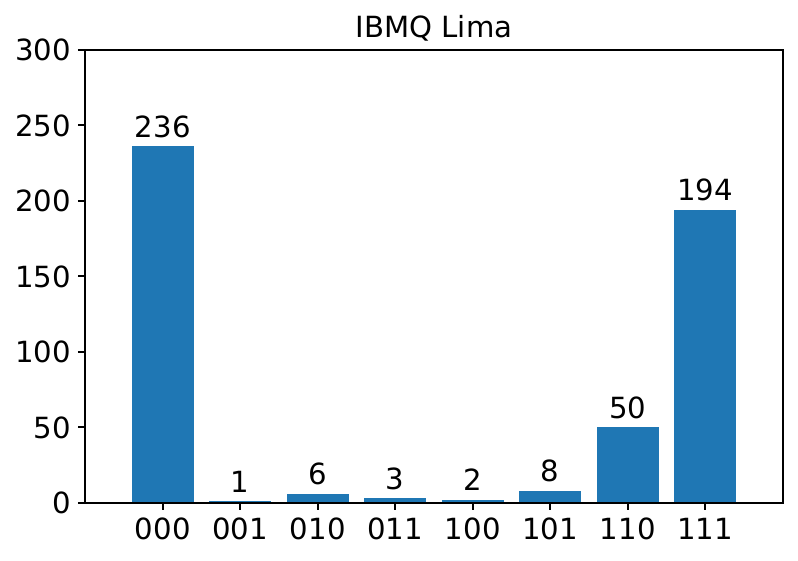} 
\endminipage

\minipage{0.49\columnwidth}
\includegraphics[width=0.9\columnwidth]{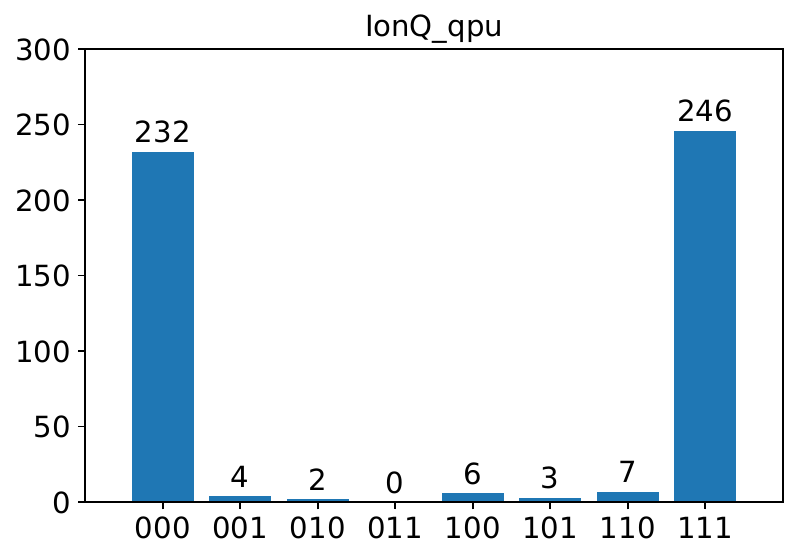} 
\endminipage\hfill
\minipage{0.49\columnwidth}
\includegraphics[width=0.9\columnwidth]{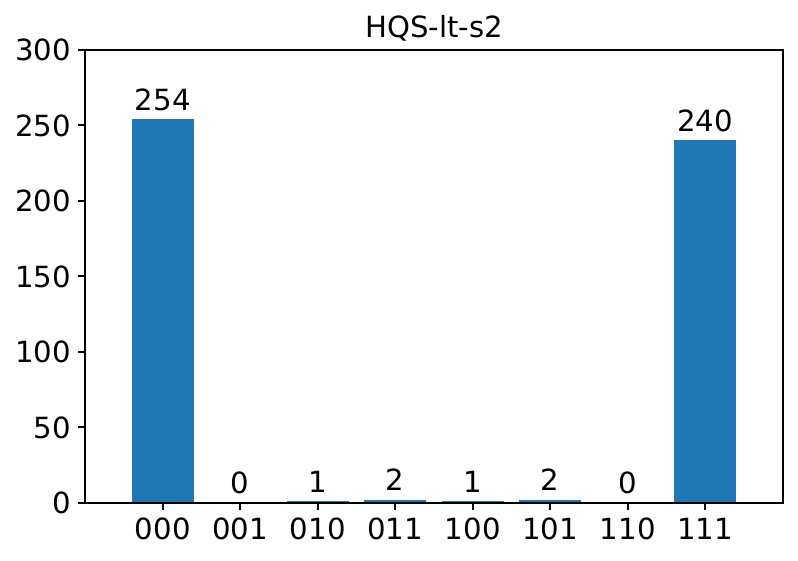} 
\endminipage
\caption{NISQ device noise illustrated by testing a 3-qubit GHZ-state circuit (top-left), showing the ground truth (top-right), Rigetti Aspen-M3 result (center-left), IBMQ-Lima (center-right), IonQ QPU (bottom-left), and Quantinuum HQS-2 result (bottom-right), using 500 induction shots.}
\label{fig:nisq_noise}
\end{figure}

Figure~\ref{fig:nisq_noise} illustrates the impact of noise on some contemporary NISQ devices through our profiling. The figure shows the sampling results of a fully-entangled quantum state by applying a 3-qubit GHZ-state circuit with 500 sampling shots. The figure displays the ground truth assuming perfect execution and the actual noisy results on Rigetti Aspen-M3, IBMQ Lima, IonQ QPU, and Quantinuum HQS-2. Within these, Aspen-M3 and Lima are based on superconducting, while QPU and HQS-2 are trapped-ion devices. As can be seen, all of them exhibit non-trivial errors in their measurements, indicating an imperfect quantum state or and/or an imperfect measurement projection due to various sources of noise still not fully understood by the community.

It is inefficient, even infeasible, to directly inspect the noise profile of a quantum device via sorely on experimental measurements, as the process tomography is notoriously costly when the device size is large. Consequently, classical computers remain a useful tool for numerically simulating quantum systems before QEC, as they allow for the investigation of NISQ device behavior with respect to noise. This is because: (1) Hypotheses regarding specific noise effects, described as noise models, as well as the corresponding QEM or QEC techniques, can be tested and validated on simulators of specific devices. (2) Feedback can be provided to quantum architects and designers regarding a particular design choice in the design space, such as selection of basis gates and topologies. (3) Critical algorithms that cannot be tested on current NISQ devices, such as quantum phase estimation (QPE) incorporating very deep circuits far beyond coherence time, can only be validated via classical simulation at present.

Efficiently simulating a quantum system numerically with noise on classical HPCs is a challenging task due to several factors. \underline{Firstly}, most existing simulators are based on functional simulation using state-vector or tensor-network representations, whereas noisy simulation requires density matrix simulation with specific noise models that take physical parameters into consideration, making the simulation approach much more complicated. \underline{Secondly}, density matrix simulation is computationally expensive, scaling in $4^n$ instead of $2^n$ for functional simulation based on state-vector representation, where $n$ is the number of qubits being simulated. For example, only to simulate the 27 qubit IBMQ Toronto device would require 256 petabytes of memory storage to hold the quantum state in density matrix form. During the simulation, each single qubit gate update requires half of the coefficients to be updated according to the gate logic, leading to significant pressure over the double-precision computation bandwidth. The computation demand can be even more extraordinary for the deep circuits of practical applications (e.g.,\cite{reiher2017elucidating}). \underline{Thirdly}, communication, synchronization, and data movement overheads pose additional challenges when scaling. As a single GPU or node cannot hold the entire wave-function due to the high memory demand, distribution of the wave-function in an HPC cluster results in significant inter-GPU communication and synchronization. As CPUs are traditionally responsible for inter-node communication, additional CPU-GPU communication and synchronization is required, further limiting performance at scale. 

This paper presents TANQ-Sim, a full-scale density matrix quantum numerical simulator designed for GPU clusters. TANQ-Sim is capable of simulating deep circuits incorporating noise generated from both coherent and non-coherent noise models. To address the significant computational cost associated with density matrix simulation, we develop a novel simulation approach based on C1 and C2 gates that represent arbitrarily configured 1- and 2-qubit gates, respectively. The C2 gate is designed to leverage the double-precision tensorcores of Ampere and Hopper GPUs, making this the first practical application of double-precision tensorcores for non-AI/ML workloads. To address deep circuits, we propose density matrix simulation gate fusion for C1 and C2 gates. To enable scalable simulation on multi-GPU clusters, we rely on the GPU-side intra-kernel communication library NVSHMEM to build a partitioned shared memory space for unified access to the stored wave-function. Instead of practicing computation-communication overlapping through fine-grained communication~\cite{li2021sv, wang2022empowering}, we found that aggregating and delegating communication to a header thread significantly improves inter-node communication performance over both Slingshot and InfiniBand network fabrics. We also design an orchestrated synchronization approach and an all-measure gate function to further improve performance. Evaluations using benchmark circuits on the NERSC Perlmutter supercomputer demonstrate the functionality, performance, and scalability of TANQ-Sim. Additionally, we use four case studies to show the practical applications of TANQ-Sim, including teleportation, distillation and Ising simulation to show the functionality of noisy simulation.

\section{Background}
In this section, we provide a brief overview of density-matrix based quantum simulation, the Ampere and Hopper double precision tensorcores, and the NVSHMEM communication library for GPU-side intra-kernel communication.

\subsection{Density Matrix Simulation}
Among the various simulation approaches, state vector based simulation is most widely used functional simulation method for dealing with pure quantum states or logical qubits. However, when dealing with a mixed system where noise is present, a pure state cannot provide sufficient information about the system. In this case, a mixed state corresponds to a statistical ensemble, or probabilistic mixture of pure states, can be used to describe the condition where a system is entangled with another system, such as the environment from which the noise is imposed. A mixed state is represented by a density matrix or a density operator, which is defined by choosing the basis in the underlying space. The density matrix is given by:
\begin{equation*}
\rho=\sum{p_s\ket{\psi_s}\bra{\psi_s}}
\end{equation*}
where $\rho$ represents the fraction of the ensemble of each pure state and is generally an unknown real value. A density matrix contains all the information of a quantum system, allowing the calculation of the probabilities of the outcomes of any measurement performed.

Compared to state-vector simulation, the density operator requires the conservation of $4^n$ coefficients, where $n$ is the number of eigenstates for each pure state, i.e., the number of qubits. Therefore, the memory cost of a density matrix simulation is $2^n$ times that of a pure state simulation using state-vector. The system also evolves according to the operator or gate sequence. When a particular gate is applied, the density matrix evolves as follows:
\begin{equation}
\rho(t)=G(t)\rho(0)G(t)^{\dagger}
 \label{eq:onegate}
\end{equation}
where $G(t)$ is the gate at time $t$, and $G(t)^\dagger$ is its adjoint. The evolution of a density matrix is more complicated than that of a state vector, considering the size of $4^n$ and the extra adjoint operator per gate. Although $G(t)$ by itself is a unitary operator, it becomes a general matrix and is not necessarily unitary in the presence of noise. Depending on the noise model, the evolution can become an ensemble of evolutions of corresponding channels, each described by a Kraus operator.

The noisy density matrix quantum numerical simulation is to compute $\rho_\text{out}$ for a n-qubit quantum system, \emph{density matrix} quantum circuit simulation is to compute $\rho_\text{out}$ for a n-qubit quantum register, given initial state $\rho_\text{in}$ and $m$ non-unitary transformations $G_0$, $G_1$, $\dots$, $G_{m-1}$:
\begin{equation}
 \rho_{\text{out}} = G_{m-1}\cdots (G_1 (G_0 \rho_{\text{in}} G^\dag_0) G^\dag_1)\cdots G^\dag_{m-1}
 \label{eq:dmsim}
\end{equation}
where $G$ and $\rho$ are $2^n\times 2^n$ matrices. Due to noise, $G$ is not necessarily a unitary matrix. $G^\dag$ is the adjoint of $G$ verifying $G^\dag=(G^*)^T$. As real quantum devices typically use 1-qubit or 2-qubit gates representing as $4\times4$ or $16\times16$ matrices, to obtain the matrix $G_i$ with a $2^n\times2^n$ size, Kronecker product or tensor product is used with the identity matrix $I$ for the other qubits.

\subsection{GPU Double Precision Tensorcores (DPTUs)}

Tensorcores have been introduced to NVIDIA GPUs since the Volta architecture to accelerate Fused Matrix Multiplication Accumulation (MMA) operations, which are fundamental operations in AI/ML algorithms. The first generation of Volta tensorcores only supports FP16 MMAs, while the second generation Turing GPUs additionally support INT8, INT4, and INT1. The third-generation Ampere tensorcores further support structured sparse matrix multiplication, the new AI/ML floating-point (BF16), the tensor floating-point (TF32), and the IEEE754 double-precision floating-point (FP64). The latest Hopper Tensorcores further support FP8. While CUDA cores perform floating-point computation at the thread granularity, tensorcores compute a small block of matrix multiplication at the granularity of a warp. Specifically, each Volta or Turing Streaming Multiprocessor (SM) contains 8 tensorcores, and each tensorcore performs a 4$\times$4$\times$4 MMA operation. In contrast, each Ampere or Hopper SM has 4 tensorcores, but each tensorcore can perform a larger 8$\times$4$\times$8 MMA operation. Previous works such as \cite{markidis2018nvidia, jia2018dissecting, sun2022dissecting} have provided detailed characterizations of the GPU tensorcores with respect to various data types, except for FP64, for which we did not find any relevant existing work.

We are particularly interested in FP64 MMA. The Ampere whitepaper~\cite{a100wp} indicate that the tensorcore theoretical FP64 throughput is 19.5 TFLOPS, which is double the 9.7 TFLOPS achieved through the ordinary CUDA cores. In Hopper, these numbers boost to 66.9 and 33.5 TFLOPS~\cite{h100wp}. A single FP64 MMA instruction can replace 8 double-precision fused-multiply-add (FMA) instructions, thereby avoiding overhead from instruction fetch and scheduling, and reducing the read bandwidth pressure over registers and shared memory. Each A100 SM can perform 64 FP64 FMA operations per cycle, with each tensorcore capable of performing 8$\times$8$\times$4 FP64 MMA. In this work, we plan to investigate the use of these new double-precision tensorcores (DPTUs) to accelerate density-matrix quantum system simulation.

\subsection{NVSHMEM for GPU-side Communication}

NVSHMEM is the state-of-the-art parallel programming interface following the \emph{Partitioned Global Address Space} (PGAS) model that provides intra-node and inter-node GPU-to-GPU communication capability for GPU clusters. It builds a global memory address space for large data that distributes its partitions across multiple GPUs, and access through fine-grained GPU-initiated peer-to-peer or collective operations. NVSHMEM integrates the capabilities from GPUDirect-RDMA \cite{potluri2013efficient} and NCCL \cite{jeaugey2017nccl}. A preliminary assessment of NVSHMEM can be found in \cite{hsu2020initial}.
  
NVSHMEM provides a means for GPU-side communication, enabling a new GPU programming paradigm in which fine-grained computation and communication can overlap with each other, thus hiding the long latency of remote communication. This is in contrast to common approaches that involve moving data to CPU buffers and relying on MPI for inter-node communication. However, as shown by a recent work \cite{wang2023mgg} using NVSHMEM to accelerate graph neural networks (GNN), NVSHMEM may not necessarily bring performance advantages, and can even lead to a 23\% performance drop when compared to a unified-virtual memory (UVM) approach. In \cite{wang2023mgg}, Wang et al. designed a software pipeline to balance local computation, memory access, and remote communication for irregular sparse computation. In our work, we propose a different approach for our simulation task.

\section{TANQ Simulation Methodology}
In this section, we describe the design of TANQ-Sim. We first present our approach of formulating the density-matrix simulation using the C1 and C2 gates. Next, we show how the C2 gate is mapped to the double-precision tensorcores. We then scale our design to multi-node, multi-GPU systems through NVSHMEM. After that, we present the design and optimization of the measurement and reset gates, which are significantly different from ordinary gates. Finally, we describe the noise models for our simulator.

\subsection{TANQ Formulation with C1 and C2 Gates}

We present our approach of density-matrix simulation. For each gate, the computation and updates to the density matrix $\rho$ follows Equation~\ref{eq:onegate}. Let us first ignore the noise and define $vec(X)$ to be column-stacking of the matrix $X$ and $unvec()$ to be the inverse operation, verifying $unvec(vec(X))=X$. We show that an arbitrary operator $G$ (not necessarily unitary) applying on qubit $q$ of a density matrix $\rho$ of size $2^n\times2^n$, expressed as $\hat{\rho}=G_q\rho G_q^{\dag}$, is equivalent to $unvec(G_{N+q}^{*}G_{q}\,vec(\rho))$:
\begin{equation}
\hat{\rho}=G_q\rho G_q^{\dag} = unvec(G_{N+q}^{*}G_{q}\,vec(\rho)) 
\end{equation}

With the linear algebra theorem stating that column-stacking $vec(ABC) = (C^{T} \otimes A) vec(B)$, $\hat{\rho}$ can be expressed as a superoperator:
\begin{align*}
vec(\hat{\rho})=vec(G_q\rho G_q^{\dag}) =\, & ((G_q^{\dag})^T \otimes G_q)vec(\rho) \\ =\, & (G_q^{*} \otimes G_q)vec(\rho)
\end{align*}
where $\otimes$ represents Kronecker product or tensor product. Consequently, we have:
\begin{align}
\hat{\rho}=unvec((G_q^{*} \otimes G_q)vec(\rho))
\end{align}
When considering noise, $G$ can be expressed as a Kraus operator, the noisy quantum channel $\xi$ is given by a set of $m$ matrices: $[K_0, K_1,\dots K_{m-1}]$ such that the evolution of the density matrix $\rho$ becomes:
\begin{align*}
\xi(\rho) = \sum_{i=0}^{m-1} K_i \rho K_i^\dagger
\end{align*}
Using the superoperator representation, we have
\begin{align*}
\xi(\rho) = unvec((\sum_{i=0}^{m-1}(K_i^{*} \otimes K_i))vec(\rho))
\end{align*}
As a result, if we store $\rho$ in column-major (i.e., $vec(\rho)$) and generate the superoperator $S=\sum_{i=0}^{m-1}(K_i^{*} \otimes K_i))$ for each noisy gate offline, the simulation is essentially to compute:
\begin{align}
\ket{\rho(t)} = S\ket{\rho(t-1)}
\label{eq:sp}
\end{align}
where $S$ is a $4^n\times4^n$ matrix, and $\ket{\rho}$ is a $4^n$ vector by column-stacking the original density matrix $\rho$. 

However, directly multiplying a $4^n\times 4^n$ matrix and a $4^n$ vector just for every single gate is too computationally expensive. Similar to the case that a $2^n\times 2^n$ matrix $G$ is formed by Kronecker product between $2\times 2$ matrix $G_q$ at qubit $q$ with identify matrix $I$ for the remaining qubits other than $q$, it can be shown that the Equation~\ref{eq:sp} can be achieved the following update approach to $\ket{\rho}$:
\begin{equation}
\begin{bmatrix}
C_{s_i} \\
C_{s_i+2^q} \\
C_{s_i+2^{q+n}} \\
C_{s_i+2^q+2^{q+n}}
\end{bmatrix}
\to
S_{4\times4}(G) \begin{bmatrix}
C_{s_i} \\
C_{s_i+2^q} \\
C_{s_i+2^{q+n}} \\
C_{s_i+2^q+2^{q+n}}
\end{bmatrix}
\end{equation}
where $s_i=\lfloor \lfloor i/{2^q}\rfloor /2^{n-1} \rfloor  2^{q+n+1} + (\lfloor i/{2^q}\rfloor \% 2^{n-1})2^{q+1} +  (i \% 2^q)$ for every integer $i\in[0,4^{n-1}-1]$. $C$ is the coefficient of $\ket{\rho}$. $S_{4\times4}(G)$ is the superoperator of the noisy gate $G$. This operation essentially performs a number of $4^{n-1}$'s $[4,4]\times[4]$ matrix-vector multiplications. Given the $4^{n-1}$ MVs are independent of each other, we can stack these vectors as columns of a new matrix, so the operation becomes a $[4,4]\times[4,4^{n-1}]$ matrix-matrix multiplication, where each element of $[4,4^{n-1}]$ has to be computed dynamically at runtime. This is in some sense analogous to the \emph{im2col()} kernel of converting a 2D convolution into a matrix-multiplication. The scenario of 2-qubit gate is similar, which is to apply $S_{16\times16}(G)$ to $\ket{\rho}$. This can be seen as a $[16,16]\times[16,4^{n-2}]$ matrix-matrix multiplication. Again, each element of the $[16,4^{n-2}]$ matrix has to be generated at runtime based on qubit positions.   

\subsection{C2 Gate Mapping to Tensorcores}

As mentioned, every physical quantum device supports a specific set of basis gates that are considered universal, and typically consist of a collection of 1-qubit and 2-qubit gates. Figure~\ref{fig:framework} shows the basis gates for IBMQ, Rigetti, IonQ and Quantinuum devices. In Section~3.1, we discussed the formulation of density matrix simulation using general 1-qubit gate C1 and 2-qubit gate C2. The computation of a single C1 gate is decomposed into $4^{n-2}$ of $[4,4]\times[4,4]$ matrix multiplication, each following the process described by Equation~\ref{eq:sp}. This is, however, too small to be effectively accelerated by the DPTUs, as each DPTU executes an MMA instruction at the granularity of $[8,4]\times[4,8]$ (see Section~2.2). Compaction can be a costly operation due to complicated address calculation. On the other hand, the computation of a  C2 gate can be decomposed into $4^{n-3}$  $[16,16]\times[16,8]$ matrix multiplications, which can be efficiently accelerated by the DPTUs.

\begin{figure}[!t]
\begin{lstlisting}[caption=TANQ-Sim C2 gate acceleration through DPTU,captionpos='b',label={lst:c2_gate}, basicstyle={\scriptsize\ttfamily\bfseries}, numberstyle=\scriptsize, backgroundcolor = \color{tinygray},framexleftmargin=7pt]
__device__ void C2_GATE(double* dm_real,double* dm_imag){
  ... // Address calculation and fetching 
  ... // coefficients from the density matrix into   
  ... // el_real_s and el_imag_s in shared memory
  __syncwarp();
  wmma::fill_fragment(c_real_up, 0.0);
  wmma::fill_fragment(c_imag_up, 0.0);
  wmma::fill_fragment(c_real_dn, 0.0);
  wmma::fill_fragment(c_imag_dn, 0.0);
  for (unsigned c=0; c<4; c++){
    // load A from const memory
    load_matrix_sync(a_real_up, &gm_real[c*4],16);
    load_matrix_sync(a_imag_up, &gm_imag[c*4],16);
    load_matrix_sync(a_real_dn, &gm_real[16*8+c*4],16);
    load_matrix_sync(a_imag_dn, &gm_imag[16*8+c*4],16);
    // load B from shared memory
    load_matrix_sync(b_real, &el_real_s[c*4],16);
    load_matrix_sync(b_imag, &el_imag_s[c*4],16);
    // complex matrix multiply for upper part of C
    mma_sync(c_imag_up, a_real_up, b_imag, c_imag_up);
    mma_sync(c_imag_up, a_imag_up, b_real, c_imag_up);
    mma_sync(c_real_up, a_real_up, b_real, c_real_up);
    a_frag_imag_up.x[0] = -a_frag_imag_up.x[0];
    mma_sync(c_real_up, a_imag_up, b_imag, c_real_up);
    // complex matrix multiply for down part of C
    mma_sync(c_imag_dn, a_real_dn, b_imag, c_imag_dn);
    mma_sync(c_imag_dn, a_imag_dn, b_real, c_imag_dn);
    mma_sync(c_real_dn, a_real_dn, b_real, c_real_dn);
    a_imag_dn.x[0] = -a_imag_dn.x[0];
    mma_sync(c_real_dn, a_imag_dn, b_imag, c_real_dn);
  } ...
}
\end{lstlisting}
\end{figure}

To operate the tensorcores using the MMA programming interface, we use each warp to independently compute a complex matrix multiplication workload of $[16,16]\times[16,8]$. Since the MMA granularity is $[8,4]\times[4,8]$, we split the C2 gate matrix A $[16,16]$ into two parts: \texttt{up} and \texttt{dn}, each with a size of $[8,16]$. Each warp then computes the \texttt{up} part of the resulting matrix C of size $[8,8]$ through four steps, using an MMA instruction to finish a block of $[8,4]\times[4,8]$ in each step. The warp then processes the lower part. We sequentially use a warp to process both \texttt{up} and \texttt{dn} parts (than distributing to two warps), because they share the same portion of data from matrix B within each step, allowing us to reuse register and exploit register locality.

Listing~\ref{lst:c2_gate} shows part of the CUDA code using DPTUs for the C2 gate. After a complicated address calculation process, we fetch required density-matrix data into two buffer arrays in shared memory. As the address calculation process involves warp branching, a warp synchronization is needed before calling to MMA. We first reset accumulation buffer of the C matrix, and then use 4 steps (Line~11) to compute a portion of $[8,8]$ of C. Within the loop, we load A from constant memory to the registers (Line~13-16) and then load matrix B from the shared memory buffer. After that, we perform complex matrix multiplication for the \texttt{up} part (Line~21-25) and the \texttt{dn} part (Line~27-31). In this way, we map the computation logic of C2 gate to the DPTUs.

\subsection{Gate Fusion}
As mentioned, quantum circuits used in practical applications can be quite deep. To address this, we propose leveraging gate fusion to merge gates and reduce circuit depth specifically for density matrix simulation.

Our strategy is simple yet effective. We start by merging consecutive 1-qubit gates that apply to the same qubit, followed by merging consecutive 2-qubit gates that apply to the same qubit pair and order. We achieve this by traversing the circuit and searching for merging opportunities. While this strategy may seem straightforward, it can still bring significant performance benefits. This is because density-matrix simulations only allow for the use of basis gates, and a general gate like an H gate has to be decomposed into a series of basis gates, which can be merged back together after noise is applied. We will demonstrate the effectiveness of this strategy in Section~5.2.

We argue that gate fusion will not impair per-gate computation intensity. Most standard quantum gates exhibit certain sparsity, with a significant number of entries being zero. For example, the Pauli-X gate ($X=[[0,1][1,0]]$) has two entries equal to zero, while the commonly used CX gate has 12 out of 16 entries being zero. Prior state-vector based simulations, such as \cite{li2021sv}, take advantage of such sparsity through gate specialization. In their approach, each gate has a specific implementation, and only non-zero calculations are performed. However, in our case, due to the presence of noise, entries may no longer be zero and we need to compute each entries for every single gate anyway. Additionally, for C2 gates, the DPTCs have to compute each entries of the gate as well. 

In addition to the basic strategy discussed earlier, we also attempted two additional strategies: First, fusing a 2-qubit gate $C2_A(q_0,q_1)$ with a 1-qubit gate $C1_B(q_0)$ that shared the same qubit $q_0$. This was achieved by converting $C1_B(q_0)$ to a 2-qubit gate $C2_B(q_0,q_1)$ through a tensor product with an identity matrix at $q_1$, and then fusing the two 2-qubit gates $C2_A(q_0,q_1)$ and $C2_B(q_0,q_1)$ together. Second, fusing two 2-qubit gates $C2_A(q_0,q_1)$ and $C2_B(q_1,q_0)$ with exchanged qubit order, by applying a SWAP gate $SWAP(q_0,q_1)$ before and after $C2_B(q_1,q_0)$: $SWAP(q_0,q_1)C2_B(q_0,q_1)SWAP(q_0,$ $q_1)$, and then merging the four 2-qubit gates into a single 2-qubit gate. Despite significant effort, both strategies were found to be unsuccessful due to the complexity of density-matrix computation and the presence of non-coherent noise.

\subsection{Scaling via NVSHMEM}

\noindent\textbf{Communication for scaling:} As described in Section~2.3, NVSHMEM creates a virtual global address space among all GPUs. For a GPU cluster like Perlmutter, each node incorporates 4 GPUs. We thus allocate the number of MPI instances to be equal to the total number of GPUs in the cluster, one MPI instance per GPU. We then use the unique MPI communicator to initialize the NVSHMEM runtime, enabling all GPUs to view and access to a unified global memory space.

Following the TANQ-Sim design described in Section~3.1, we represent the density matrix $\rho$ as a column-stacked vector $vec(\rho)$. Consequently, we use column-major indexing for the density matrix $\rho$ and partition it among the symmetric memory of all available GPUs. To facilitate remote access, we unify the access using the macros defined in Listing~\ref{lst:comm} for remote put and get operations. Here, \emph{m\_gpu} refers to the number of density matrix elements per GPU, and \emph{lg2\_m\_gpu} is the $\log_2$ of \emph{m\_gpu}, used for efficient address calculation through shifting operations.

\begin{figure}[!t]
\begin{lstlisting}[caption=TANQ-Sim communication and synchronization through NVSHMEM.,captionpos='b',label={lst:comm}, basicstyle={\scriptsize\ttfamily\bfseries}, numberstyle=\scriptsize, backgroundcolor = \color{tinygray},framexleftmargin=7pt]
//define the unified remote put function
#define PGAS_P(arr,i,val) nvshmem_double_p(&(arr)[(i)\
    &(m_gpu-1)], (val), ((i)>>lg2_m_gpu))
//define the unified remote get function
#define PGAS_G(arr, i) nvshmem_double_g(&(arr)[(i)\
    &(m_gpu-1)], ((i)>>lg2_m_gpu))
//define global barrier 
#define BARR                                 \
  if (threadIdx.x == 0 && blockIdx.x == 0)   \
    nvshmem_barrier_all();                   \
    grid.sync();
}
__device__ void C1_GATE(){
  ...
  // load data from remote GPU to local buffer
  if (tid == 0) nvshmem_double_get(dm_real_remote,
    dm_real, per_pe_num, pair_gpu);
  if (tid == 0) nvshmem_double_get(dm_imag_remote,
    dm_imag, per_pe_num, pair_gpu);
  grid.sync();
  ...
  //store data from local buffer to remote GPU
  grid.sync();
  if (tid == 0) nvshmem_double_put(dm_real,
    dm_real_remote, per_pe_num, pair_gpu);
  if (tid == 0) nvshmem_double_put(dm_imag,
    dm_imag_remote, per_pe_num, pair_gpu);
}
__device__ void Gate::exe_op(Simulation *sim)
{
  grid_group grid = this_grid();
  // global barrier only when remote access is performed
  if (((ctrl+n_qubits)>=lg2_m_gpu) ||
    ((qubit+n_qubits)>=lg2_m_gpu)) BARR;
  ... //call to specific gates
  grid.sync(); //ensure per-GPU local consistency
}
\end{lstlisting}
\end{figure}

In our initial design, we used fine-grained NVSHMEM remote access to fetch and update individual elements of the density matrix with respect to Equation~\ref{eq:sp}. However, as the density matrix grew in size, the large number of fine-grained remote accesses per gate soon overloaded the network fabric buffers, leading to program hang-ups. Since NVSHMEM is a relatively new programming interface, there are rare relevant materials available, especially regarding optimization. To address this issue, we explored different approaches and identified two general optimization strategies: (1) aggregating fine-grained remote access into larger chunks and using a single delegate thread to perform coarse-grained data migration, which significantly improves communication efficiency; and (2) using bidirectional communication, which can better utilize the fabric bandwidth. Our approach is depicted in Listing~\ref{lst:comm}.

\vspace{4pt}\noindent\textbf{Synchronization for scaling:} To maintain data consistency in the density matrix across consecutive gates, global synchronization is necessary after each gate. In Listing~\ref{lst:comm}, we show how we achieve global synchronization \emph{BARR} among all threads of all GPUs. First, we synchronize among the leading threads of all GPUs through NVSHMEM primitive, then we synchronize among all threads of the local GPU grid. Despite functioning well, we found the global synchronization of this approach to be extremely costly. To reduce such overhead, we have developed a logic to determine whether a gate needs to perform remote access. If so, we perform global synchronization; otherwise, we do GPU local synchronization. This ensures data consistency while minimizing synchronization overhead,  shown in Line~33-36 of Listing~\ref{lst:comm}.

\subsection{Noise Models}
In order to construct accurate noise models, we conducted a thorough investigation of various noise modeling approaches from different quantum computing providers, such as IBMQ, Google's Cirq, Rigetti's PyQuil, and Microsoft's Q\#. Although each provider's approach shared similarities, we ultimately chose to base our noise model design with respect to the Qiskit-Aer approach, due to its comprehensive documentation and the widespread adoption of IBMQ devices. Specifically, TANQ-Sim realizes the following noise models:

\vspace{4pt}\noindent\textbf{Gate Depolarizing Channel Error:} The depolarizing channel error is the most commonly observed type of qubit error in quantum computing, characterized by the probability of a qubit's state being flipped due to interactions with its environment. This type of coherent error is referred to as \emph{depolarizing} because it can cause the qubit's state to become mixed or depolarized, making it difficult to extract information from the qubit. However, depolarizing channel errors can be corrected through the use of quantum error correction (QEC) codes. In TANQ-Sim, we model both 1-qubit and 2-qubit depolarizing errors to simulate their effects on the execution of quantum circuits.

\vspace{4pt}\noindent\textbf{Gate Thermal Relaxation Error:} Thermal relaxation is a phenomenon in which a quantum system loses energy to its environment due to thermal fluctuations, leading to randomization of the system. In TANQ-Sim, we model both amplitude damping and phase damping using decay probabilities as inputs. Amplitude damping occurs when the quantum system loses energy to its environment, resulting in a reduction in the amplitude of the qubit state. Phase damping occurs when the system loses its phase information due to interactions with the environment. Both types of damping are non-coherent errors that can be challenging to correct directly through QEC. In TANQ-Sim, we model both 1-qubit and 2-qubit thermal relaxation errors to assess their impact on the execution of quantum circuits.

In TANQ-Sim, we generate all noisy gate matrices as Kraus operators, following their definitions, and then convert them to Liouville superoperators for density-matrix simulation. To facilitate this process, we have developed a script that fetches the necessary device information from, for example, an IBMQ backend object, including T1, T2, frequency, readout length, readout error, gate length, gate error for all basis gates, and CX coupling or topology. This allows us to accurately model the noise and errors present in quantum computing hardware and evaluate the performance of quantum circuits under realistic conditions.

\vspace{4pt}\noindent\textbf{Measurement Error:} Qubit measurement error is a type of error that occurs during the measurement of a qubit in a quantum computing system. It can be caused by imperfections in the measurement apparatus, environmental noise, or interactions between the measured qubit and other parts of the quantum system. The measurement error can be further divided into prob\_meas0\_prep1 and prob\_meas1\_prep0, which respectively represent the probabilities of preparing a given computational basis state but measuring the orthogonal state (P(0|1) and P(1|0)). The measurement error is taken as the average of these two values. It should be noted that prob\_meas0\_prep1 is typically the dominant factor in read-out error due to relaxation during the measurement process.

\begin{table}[!t]
\centering\small
\caption{Consistency of noise channels in TANQ-Sim and IBMQ device backend noise models. The metrics are the average norm difference between gate-noise operators over all qubits or couplings, and the range of $p$-values for measurement samplings over all qubits.}
\begin{tabular}{|c|c|c|c|c|c|c|}
\hline
\textbf{Backend} & \textbf{Qubits} & \textbf{ID} & \textbf{SX} & \textbf{X} & \textbf{CX} & \textbf{Meas.}  \\ \hline

Lima &5 &5.8E-16 &5.8E-16 &5.8E-16 &1.8E-15 &$[0.56,1]$ \\ \hline
Cairo &27 &5.9E-16 &5.9E-16 &5.9E-16 &2.2E-15 &$[0.31,1]$ \\ \hline
Mumbai &27 &6.1E-16 &6.1E-16 &6.1E-16 &2.5E-15 &$[0.43,1]$ \\ \hline
Brooklyn &65 &5.8E-16 &5.8E-16 &5.8E-16 &2.5E-15  &$[0.50,1]$  \\ \hline
\end{tabular}
\label{tab:noise_model_eval}
\end{table}

\vspace{4pt}\noindent\textbf{Validation of Gate Error Models:} We validate our gate noise models by comparing the superoperator matrices of noise channels generated by TANQ-Sim and Qiskit-Aer using the same input parameters. The validation process consists of two procedures: (I) For a given noise channel $\mathcal{E}$ and input parameter(s) $\alpha \in D$, we discretize $D$ and iterate over all discretizations to produce the superoperator matrices of channel $\mathcal{E}$, $\mathcal{M}_{TANQ}$ in TANQ-Sim and $\mathcal{M}_{Aer}$ in Qiskit-Aer for every choice of $\alpha$. We then check if the norm of the difference between $\mathcal{M}_{TANQ}$ and $\mathcal{M}_{Aer}$ is less than a specified tolerance, $\epsilon \approx 10^{-14}$. (II) For a selected IBMQ quantum device, we read the device's calibration data and generated the superoperator matrices that describe the noise present in each of the device's basis gates: \texttt{ID}, \texttt{SX}, \texttt{X}, and \texttt{CX}. We then compared the superoperator matrices generated by TANQ-Sim with those provided by Qiskit-Aer using the NoiseModel.from\_backend() method, following the same procedure as in the previous step. It should be noted that since IBMQ implements the \texttt{RZ} gate as a classical operation, no noise is essentially present in the \texttt{RZ} gates.

We report the average norm difference over all qubits in the second procedure for four different IBMQ devices in Table~\ref{tab:noise_model_eval}. The results show that our noise models accurately match the noise simulation settings for each of the listed IBMQ devices, indicating the robustness and reliability of our modeling approach.

\vspace{4pt}\noindent\textbf{Validation of Measurement Error Models:} Direct comparison of measurement error channels is not feasible due to different implementations. To validate our measurement error models, we performed two similar procedures to the previous gate-noise validation, with the following modifications: For a given density matrix $\rho$ and the parameters {P(1|0), P(0|1)}, we computed the noisy probability of measuring 0, $P_{TANQ}$, in TANQ-Sim. We then estimated 30 samples of the same probability, ${P_{1,Aer}, \dots, P_{30, Aer}}$, using identical input parameters (either from discretization over parameter domain or real-backend simulation) in the Qiskit-Aer density-matrix simulator by measuring $\rho$ for $30 \cdot 1000$ shots. Under the null hypothesis $H_0: P_{TANQ} = \text{mean}(\{P_{k,Aer}\})$, we performed one-sample $t$-tests and reported the resulting $p$-value.

\begin{figure*}[!t]
\centering
\includegraphics[width=1.7\columnwidth]{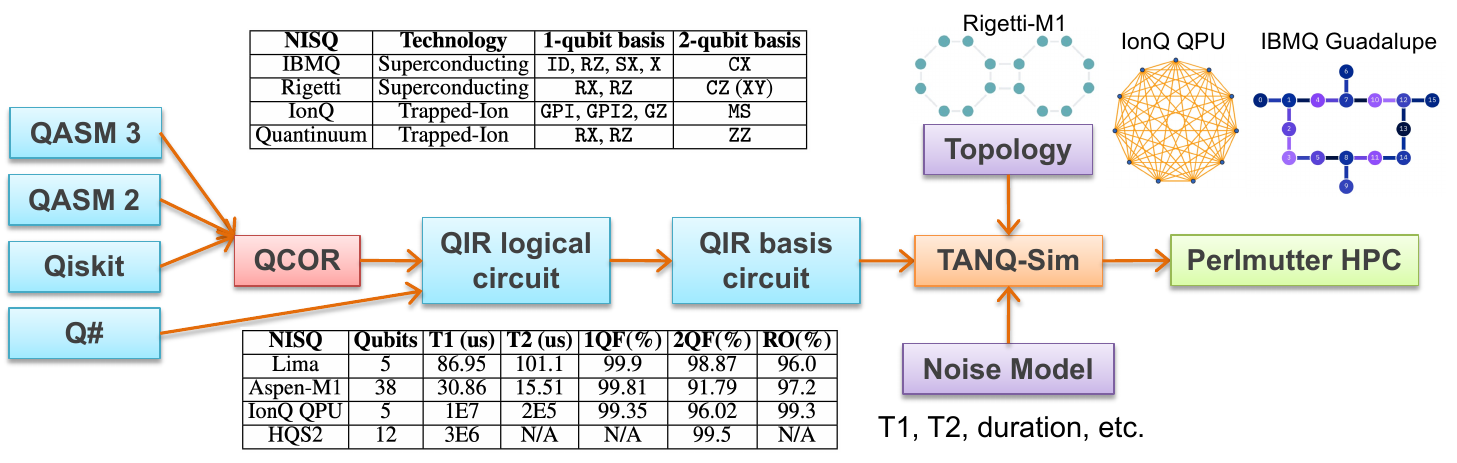}
\caption{TANQ-Sim infrastructure. 1QF refers to the average 1-qubit fidelity. 2QF refers to the average 2-qubit fidelity. RO stands for average read-out fidelity.}
\label{fig:framework}
\end{figure*}

The range of $p$-values over all qubits in each simulated backend noise model is presented in Table~\ref{tab:noise_model_eval}. Notably, all $p$-values are very large, indicating that there is no evidence to reject the null hypothesis. This implies that the difference between our measurement error model and that of IBMQ is not statistically significant. It should be noted that the comparisons in Table~\ref{tab:noise_model_eval} were performed using Qiskit simulated backend-noise models, not the real IBMQ devices.

\section{TANQ-Sim Framework}

The overall infrastructure of TANQ-Sim is illustrated in Figure~\ref{fig:framework}. TANQ-Sim supports various front-ends, including Q\#, QASM2, QASM3, and Qiskit through the Microsoft \emph{Quantum Intermediate Representation} (QIR). QIR is a language-agnostic, low-level, LLVM-based intermediate representation that is designed to enable efficient and interoperable execution of quantum programs across different quantum hardware and software platforms. It allows quantum programs to be compiled into a common representation that can be optimized and translated into executable code for a specific backend, such as TANQ-Sim in this case. Q\# can dump QIR with a special emit instruction added to the Q\# project XML file. The other front-ends are compiled into QIR via QCOR \cite{mintz2020qcor}.

\begin{table}[!t]
\centering\footnotesize
\caption{Standard QIR runtime gates. "MC" refer to multi-controlled.}
\begin{tabular}{|c|l|c|l|}
\hline
\textbf{Gates} & \textbf{Meaning} & \textbf{Gates} & \textbf{Meaning}  \\ \hline

\texttt{X} & Pauli-X bit flip &  \texttt{ControlledX} & MC \texttt{X} \\\hline
\texttt{Y} & Pauli-Y bit/phase flip &  \texttt{ControlledY} & MC \texttt{Y}  \\\hline
\texttt{Z} & Pauli-Z phase flip  &  \texttt{ControlledZ} & MC \texttt{Z}  \\\hline
\texttt{H} & Hadamard  &  \texttt{ControlledH} & MC \texttt{H} \\\hline
\texttt{S} & sqrt(Z) phase &  \texttt{ControlledS} & MC S   \\\hline
\texttt{T} & sqrt(S) phase  &  \texttt{ControlledT} & MC T  \\\hline
\texttt{R} & Unified rotation gate  & \texttt{ControlledR} & MC R \\\hline
\texttt{Exp} & Exponential of Pauli &  \texttt{ControlledExp} & MC Exp \\\hline
\texttt{AdjointS} & Adjoint S & \texttt{ControlledAdjointS} & MC AdjointS \\\hline
\texttt{AdjointT} & Adjoint T  & \texttt{ControlledAdjointT} & MC AdjointT \\\hline

\end{tabular}
\label{tab:qir_gates}
\end{table}

The QIR logical circuit in Figure~\ref{fig:framework} is expressed using standard QIR runtime gates listed in Table~\ref{tab:qir_gates}. These gates are expected to be supported by a backend by default. However, TANQ-Sim simulates quantum devices that only support a limited set of basis gates, such as \texttt{ID}, \texttt{SX}, \texttt{X}, \texttt{RZ}, and \texttt{CX} for IBMQ, as shown in Figure~\ref{fig:framework}. To address this limitation, we developed a module in the QIR runtime that can decompose standard QIR gates into particular basis gates, depending on the backend, such as IBMQ, Rigetti, IonQ, and Quantinuum. Using this transpilation module, we obtain a QIR basis circuit composed only of basis gates. We then inject noise into the basis gates using hardware calibration data, such as T1, T2, and gate duration, as described in Section3.5, and perform mapping with respect to topology or coupling constraints. The obtained circuit, consisting solely of C1 and C2 gates after gate fusion described in Section~3.3, is executed distributively through NVSHMEM, as described in Section~3.4, on GPU-accelerated HPC clusters, such as NERSC Perlmutter.

\begin{figure}[!t]
\centering
\includegraphics[width=0.8\columnwidth]{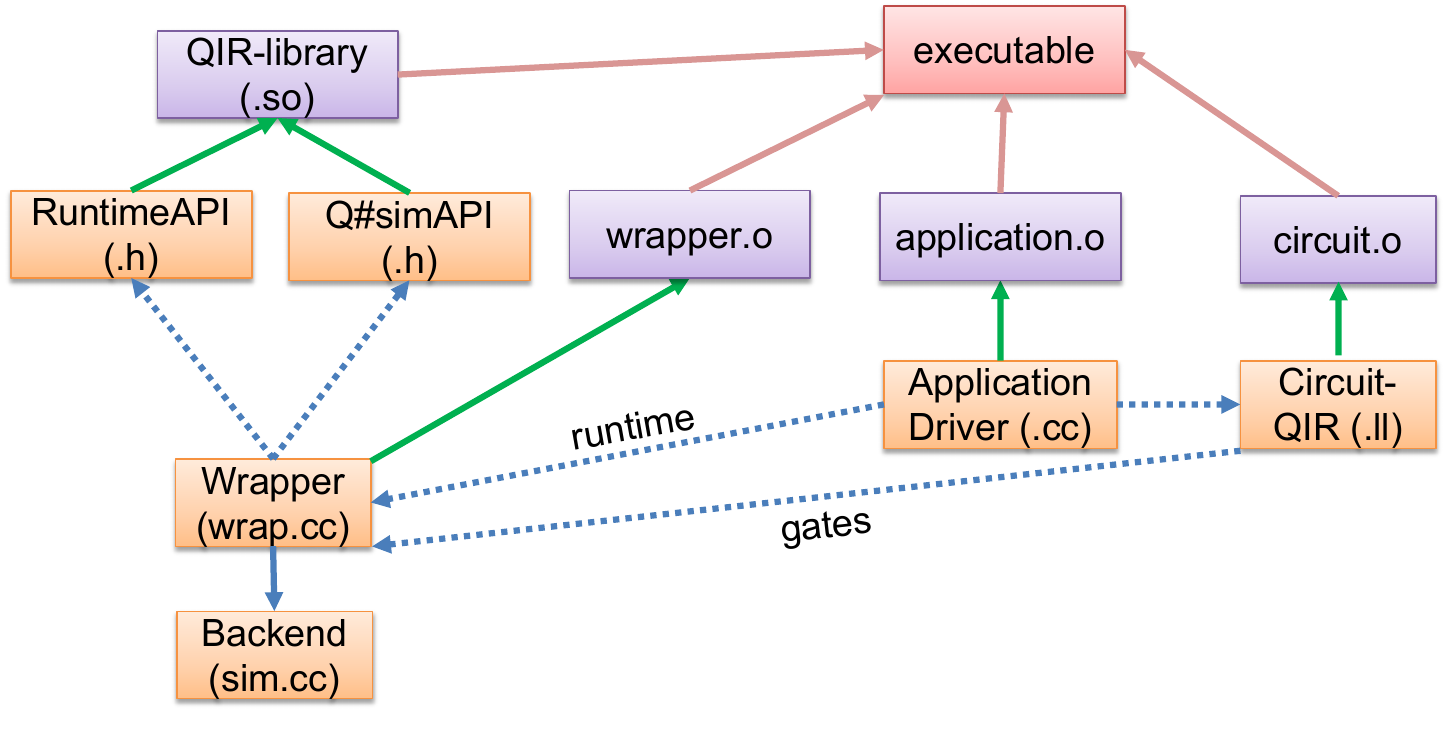}
\caption{The integration of TANQ-Sim with QIR.}
\label{fig:qir_wrap}
\end{figure}

To support QIR in TANQ-Sim, we developed a wrapper, as illustrated in Figure~\ref{fig:qir_wrap}. The wrapper is derived from the Q\# Sim-API and Runtime-API. The Sim-API header file declares the standard gates listed in Table~\ref{tab:qir_gates}, while the Runtime-API header file declares the runtime functions to be realized, such as qubit allocation and release, result release, and so on. During compilation, the QIR file of the circuit is compiled into an object file using LLVM Clang. The application code, which calls to the QIR file for circuit invocation, and the wrapper for runtime management are also compiled into an object file. These object files, together with the wrapper object, the TANQ-Sim object, the QIR library, the NVSHMEM library, and other necessary libraries, can be linked to obtain the final executable. Figure~\ref{fig:qir_wrap} and the description provided here can be useful for other devices or simulator backends to support QIR, or other front-ends through QCOR and QIR.

\section{Evaluation}

This section presents the evaluation of TANQ-Sim. We begin by describing the HPC system we used, followed by the benchmark circuits and evaluation results. Finally, we present four case studies to show the usability of TANQ-Sim. 

\subsection{Perlmutter System Settings}
We use the NERSC Perlmutter HPC system for evaluation. Perlmutter is an HPE Cray EX system with both CPU and GPU partitions. It is a pre-exascale HPC system based on the HPE Cray Shasta platform. The heterogeneous system comprises 1536 GPU nodes (each contains four NVIDIA A100 40GB GPUs), 256 GPU nodes (four A100 80GB GPUs), and 3072 CPU nodes (AMD EPYC 7763) linked by HPE Slingshot 11 high-speed interconnect. Each Perlmutter node has 4 NVIDIA A100 GPUs, connected by NVLink-V3, and an AMD Milan EPYC 7763 64-core CPU. The host compiler is nvc++ 21.11. We use CUDA Runtime 11.5.

\subsection{Benchmark Results}

\begin{table}[!t]
\centering\footnotesize
\caption{Benchmark circuits from QASMBench~\cite{li2020qasmbench}. "Basis" refers to the number of basis gates.}
\begin{tabular}{|c|l|c|c|c|c|c|c|c|}
\hline
\textbf{Circuit} & \textbf{Algorithm} & \textbf{Qubits} & \textbf{Gates} & \textbf{Qubits} & \textbf{Basis} \\ \hline
Adder  & Quantum Arithmetic & 10 & 142 & 10 & 358 \\ \hline
BV  & Hidden Subgroup & 14 & 41 & 15 & 183 \\ \hline
Ising  & Quantum Simulation & 10 & 480 & 15 & 962 \\ \hline
Multiplier  & Quantum Arithmetic & 15 & 574 & 15 & 1571 \\ \hline
QF21  & Quantum Arithmetic & 15 & 311 & 15 & 801 \\ \hline
QFT  & Hidden Subgroup & 15 & 540 & 15 & 1471 \\ \hline
QPE  & Hidden Subgroup & 9 & 123 & 9 & 271 \\ \hline
SAT  & Search and Optimization & 11 & 679 & 15 & 1589 \\ \hline
SECA  & Error Correction & 11 & 216 & 15 & 430 \\ \hline
VQE  & Search and Optimization & 8 & 10808 & 8 & 32826  \\ \hline
DNN &  Machine Learning & 16 & 2016 & 16 & 8345  \\ \hline
\end{tabular}
\label{tab:basis_gate_param}
\end{table}

\begin{figure}[!t]
\centering
\includegraphics[width=0.8\columnwidth]{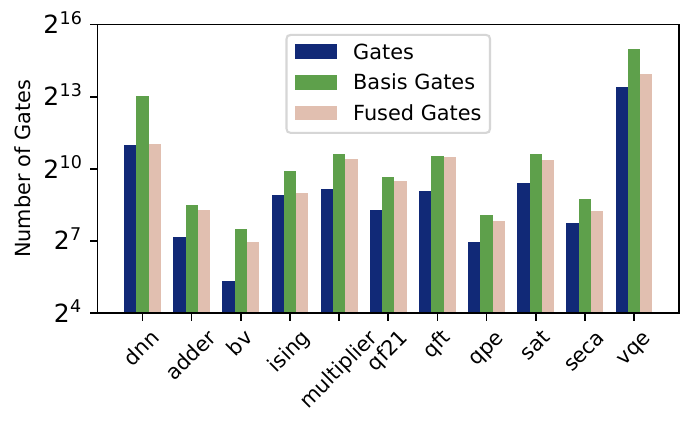}
\caption{Number of gates with transpilation and gate fusion.}
\label{fig:gate_fusion}
\end{figure}

We use the application benchmark circuits from QASMBench \cite{li2020qasmbench}, as listed in Table~\ref{tab:basis_gate_param}. We use the 16-qubit \emph{ibmq\_guadalupe} as the targeted device for transpilation and noisy simulation. The topology and basis gate set are shown in Figure~\ref{fig:framework}.

\noindent\textbf{Gate Fusion:} We first evaluate the effectiveness of the proposed gate fusion approach for density matrix simulation. Figure~\ref{fig:gate_fusion} shows the number of gates in the original QASM2 circuits, the number of basis gates after transpilation by QIR, and the number of gates after applying gate fusion described in Section~3.3. As can be seen, on average the transpilation leads to 2.8x increases in gate count, while our gate fusion technique can reduce the number back by 1.6x on average. 

\begin{figure}[!t]
\centering
\includegraphics[width=0.7\columnwidth]{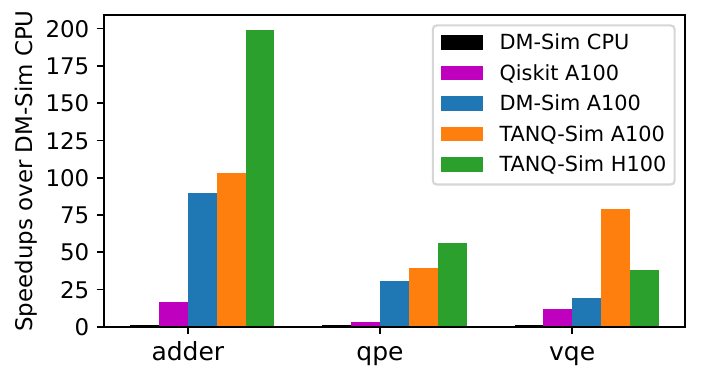}
\caption{Performance comparison with Qiskit-Aer-GPU, DM-Sim, and across GPU architectures.}
\label{fig:perf_comp}
\end{figure}

\vspace{4pt}\noindent\textbf{Performance Comparison:} We compare TANQ-Sim  with the latest Qiskit-Aer GPU simulator (Version 0.11.2) through density matrix representation using an A100 GPU of Perlmutter. Additionally, we compare TANQ-Sim with another existing density matrix simulator DM-Sim~\cite{li2020density}, using its CPU and GPU backends on Perlmutter. Furthermore, to show performance across GPU architectures, we also run TANQ-Sim on an NVIDIA H100 SXM5 GPU with Hopper architecture and tensorcores. Since the circuits with 15-qubit after transpilation (see Table~\ref{tab:basis_gate_param}) use too much GPU memory and failed in DM-Sim, we only show the results on the three circuits \texttt{adder}, \texttt{qpe}, and \texttt{vqe}. We use DM-Sim CPU backend performance as the baseline, and normalize others with respect to the baseline, shown as speedups in Figure~\ref{fig:perf_comp}. As can be seen, the TANQ-Sim shows significantly improved performance over existing density matrix simulators.   

\begin{figure}[!t]
\centering
\includegraphics[width=0.8\columnwidth]{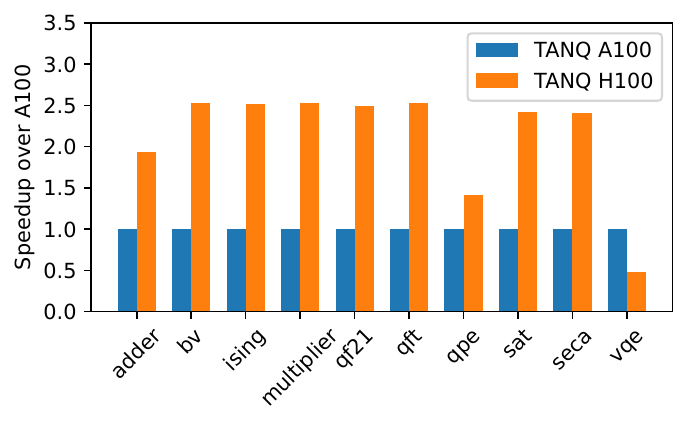}
\caption{TANQ-Sim performance on A100 and H100 GPUs.}
\label{fig:h100}
\end{figure}

Figure~\ref{fig:h100} shows the performance comparison between A100 and H100 GPUs. Overall, for TANQ-Sim, the H100 achieves on average 2.12x over A100, less than the 3x theoretical speedup~\cite{h100wp}. 

\vspace{4pt}\noindent\textbf{Performance Scaling:} As quantum circuit simulation overhead increases exponentially with the number of qubits being simulated, achieving strong scaling performance across larger numbers of compute nodes is crucial for handling larger circuit simulations.

\begin{table}[h!]
\centering\footnotesize
\caption{Larger Benchmark Circuits}
\begin{tabular}{|c|l|c|c|c|}
\hline
\textbf{Circuit} & \textbf{Algorithm} & \textbf{Qubits} & \textbf{Gates} & \textbf{Density Matrix Size} \\ \hline
DNN  & Machine Learning & 16 & 1041 & 128 GB \\ \hline
QFT  & Hidden Subgroup & 16 & 611 & 128 GB \\ \hline
QV  & Quantum Volume & 17 & 36 & 512 GB \\ \hline
QEC  & Surface Code & 17 & 54 & 512 GB \\ \hline
Adder  & Quantum Arithmetic & 18 & 277  & 2 TB  \\ \hline
SQRT  & Quantum Arithmetic & 18 & 2143 & 2 TB \\ \hline
RCS  & Random Circuit Sampling & 19 & 137 & 8 TB \\ \hline
BV  & Hidden Subgroup & 19 & 57 & 8 TB  \\ \hline
QRAM  & Quantum Memory & 20 & 308 & 32 TB \\ \hline
QAOA  & Search and Optimization & 20 & 101 & 32 TB  \\ \hline
GHZ & Entanglement Generation & 21 & 43 & 128 TB  \\ \hline
W\_State & Entanglement Generation & 21 & 103 & 128 TB \\ \hline
\end{tabular}
\label{tab:larger_circuits}
\end{table}

To evaluate this scaling, we tested larger benchmark circuits shown in Table~\ref{tab:larger_circuits} with 16 to 21 qubits. Notably, the largest benchmark circuits require 128 TB of memory. For these tests, we employed a larger device noise profile - a 30-qubit device with similar gate and decoherence error rates compared to the 16-qubit ibmq\_guadalupe device used previously. This device features all-to-all qubit connectivity and it is chosen to prevent the introduction of additional ancilla qubits during circuit transpilation.

\begin{figure}[h!]
\centering
\includegraphics[width=1\columnwidth]{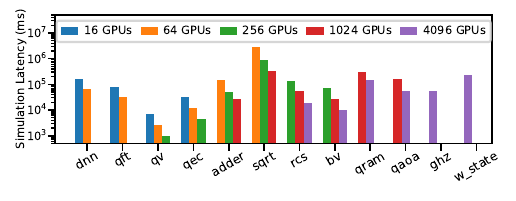}
\caption{TANQ-Sim scaling performance on Perlmutter.}
\label{fig:perlmutter_storng}
\end{figure}

Our tests are running on the Perlmutter HPC system. We use the compute node that each contains 4 A100 40GB GPUs. We scaled from 4 nodes (16 GPUs) up to 1024 nodes (4,096 GPUs). For each node count, we selected benchmarks that would neither exceed the GPU memory limits nor fall below 2GB per GPU memory usage, ensuring meaningful problem sizes for each configuration.
In our strong scaling evaluation of TANQ-Sim, Figure~\ref{fig:perlmutter_storng} demonstrates the performance scaling of simulation time across circuit benchmarks. The results show a great scaling of TANQ-Sim with a median parallel efficiency of 67\%.

\begin{figure}[h!]
\centering
\includegraphics[width=1\columnwidth]{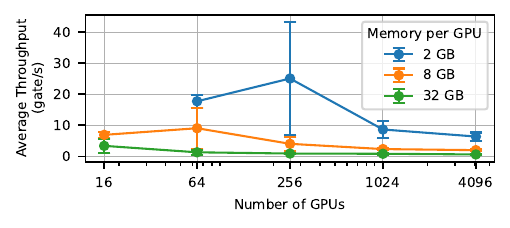}
\caption{Weak scaling of TANQ-Sim on Perlmutter.}
\label{fig:perlmutter_weak}
\end{figure}

For weak scaling performance, we used per-GPU memory usage as a metric for problem complexity and measured simulation throughput (gates simulated per second) against the number of nodes (GPUs). Figure~\ref{fig:perlmutter_weak} presents these results. We observed an anomalous high throughput data point for the 2GB per GPU configuration at 256 GPUs, corresponding to the 17-qubit Quantum Volume circuit. This outlier stems from qv\_n17's unusual gate composition, where 80\% are 1-qubit gates, which typically require less computational resources than 2-qubit gates. Despite this anomaly, TANQ-Sim demonstrates excellent weak scaling performance, maintaining consistent throughput for fixed per-GPU problem sizes as GPU count increases.

\subsection{Case Study-1: Teleportation}

\begin{figure*}[!h]
     \centering
     \begin{subfigure}[h]{0.31\textwidth}
         \centering
         \includegraphics[width=1.0\textwidth]{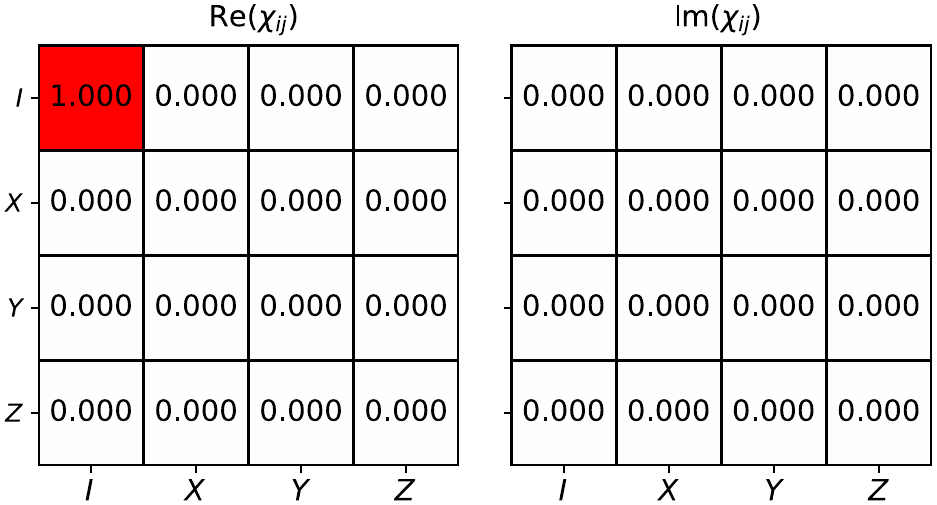}
         \caption{Teleportation with ideal gates.}
         \label{fig:tele_noiseless}
     \end{subfigure}
     \hfill
     \begin{subfigure}[h]{0.31\textwidth}
         \centering
         \includegraphics[width=1.0 \textwidth]{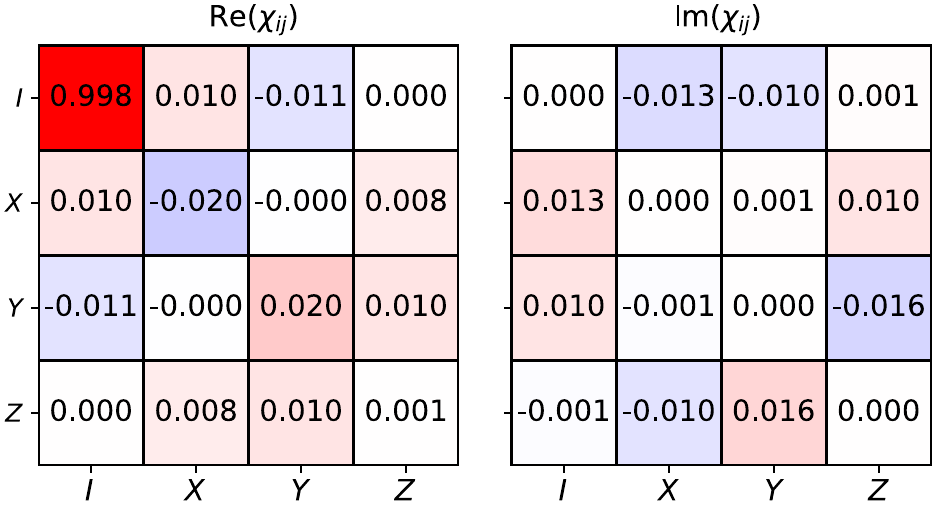}
         \caption{Teleportation with minor noise.}
         \label{fig:tele_toronto}
     \end{subfigure}
     \hfill
      \begin{subfigure}[h]{0.35\textwidth}
         \centering
         \includegraphics[width=1.0\textwidth]{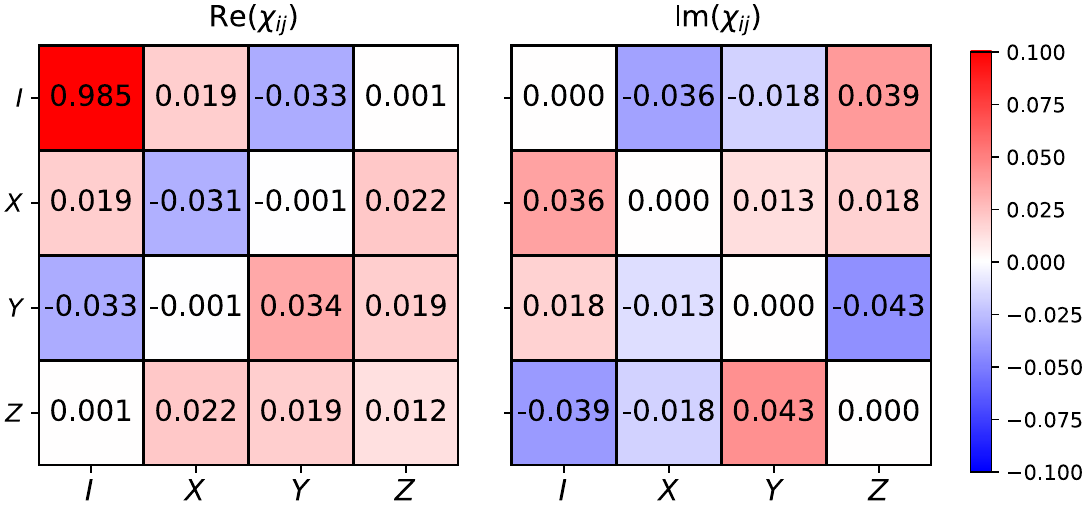}
         \caption{Teleportation with increased noise.}
         \label{fig:tele_toronto_worse}
     \end{subfigure}
        \caption{TANQ-Sim results over a teleportation simulation circuit. Matrices plotted in the upper row represent real values, and lower rows complex values of the $\chi$ matrix.  We plot the $\chi$ matrix from the channel tomography~\cite{nielsen2002quantum}. The noiseless channel is shown in (a), while we consider including gate errors, measurement errors and shot noises in (b) and (c). We use {\it ibm\_toronto} device noise model to compute the $\chi$ matrix of the teleportation channel in (b). We further turn up gate error rates of single-qubit and two-qubit gates to $0.1$, and plot the $\chi$ matrix of this teleportation channel in (c).}
        \label{fig:teleportation}
\end{figure*}

Quantum teleportation is a protocol that transfers a quantum state from one location to another without physically moving the qubits. By establishing entanglement between two quantum states in different locations and performing both measurements and conditional corrective operations, a quantum state is teleported over the entangled pair.

We simulated the 3-qubit teleportation circuit, transpiled it to the basis gate set, and simulated it using TANQ-Sim. We try to simulate the impact on teleportation quality due to noise, i.e., the deviation of the teleportation channel from the ideal channel. Figure~\ref{fig:teleportation} illustrates the $\chi$-matrix of each different channel through process tomography~\cite{nielsen2002quantum}. Using the $\chi$-matrix, the quantum channel can be represented as 
\begin{equation}
\rho_{\text{new}} = \mathcal{E}(\rho) = \sum_{i,j} \chi_{ij} \hat{\sigma}_i \rho \hat{\sigma}_j,
\end{equation}
where $\chi_{ij}$ are $\chi$ matrix elements, $\rho$ is the input qubit single-qubit density operator, and $\mathcal{E}$ represents the teleportation channel. The indices $i,j$ summation goes over $0$ to $3$ for single qubit channels, where $\hat{\sigma}_0$ is an identity operator, while $\hat{\sigma}_{1,2,3} = \hat{\sigma}_{x,y,z}$ are Pauli matrices.

We consider three cases. In Figure~\ref{fig:tele_noiseless}, we plot the $\chi$-matrix of a teleportation channel without any noise. Note that the ideal teleportation channel is an identity channel, where
\begin{equation}
    \chi_{ij} = \delta_{i,0} \delta_{j,0}.
\end{equation}
In Figure~\ref{fig:tele_toronto}, we compute the teleportation channel $\chi$-matrix with realistic noise. We use the configurations of {\it ibm\_toronto} device, including gate errors, limited decoherence time of qubits, and measurement errors. As shown in Figure~\ref{fig:tele_toronto}, the teleportation channel deviates from the ideal channel due to imperfect operations. To further show the effects of imperfections and the capability of TANQ-Sim, we manually increase the single-qubit and two-qubit gate error to $0.1$ and show the results in Figure~\ref{fig:tele_toronto_worse}. These Hinton diagrams illustrates how particular changes of a noise model parameter, potentially due to improved quantum devices, can make an impact on the fidelity of quantum teleportation, which is the fundamental process for quantum communication~\cite{ang2022architectures} and distributive quantum computing~\cite{wu2022autocomm, wu2023qucomm}.

\subsection{Case Study-2: Distillation}

\begin{figure}[!t]
    \centering
    \includegraphics[width = 0.65 \columnwidth]{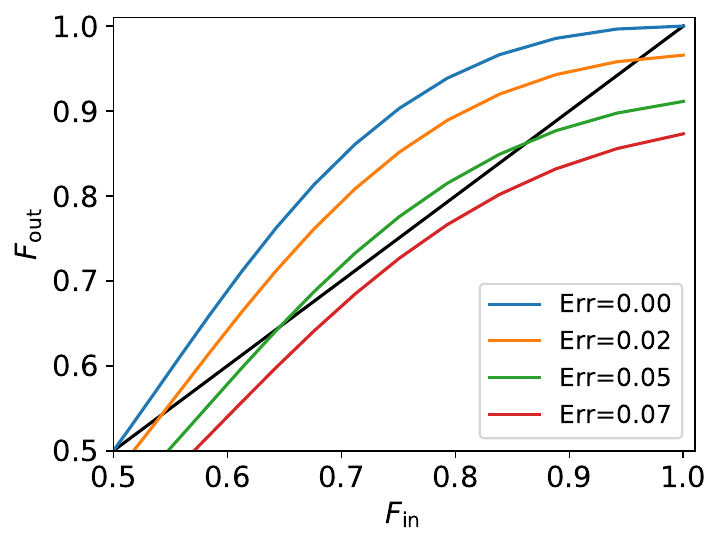}
    \caption{Purification with remote CX gate errors. Here we calculate the fidelity of the final Bell pair after two rounds of successful nested purification~\cite{deutsch1996}. The input flawed Bell pair fidelity is shown as the $x$ axis. The noiseless result is shown as the blue curves. We then gradually increase remote CX gate errors which decreases the purification results. The black line shows $F_{\text{out}} = F_{\text{in}}$.}
    \label{fig:purification}
\end{figure}

Distillation is another foundational protocol for quantum communication and distributed quantum computing~\cite{ang2022architectures}. Distillation protocols consume many noisy bell pairs to produce fewer, higher fidelity pairs. These protocols can be iteratively executed, consuming several already purified pairs to generate pairs with even better fidelity. However, noise such as decay prohibits the recursive distillation to purity, resulting in purification reducing bell pair fidelity rather than improving it.

We use TANQ-sim to explore potential density matrix base error models and their corresponding entanglement distillation performance. As shown in Figure~\ref{fig:purification}, we observe that with error rates above 0.07, distillation can no longer offer any improvements with a given input fidelity. However, as we reduce error rates, we observe wider bands of fidelities that can gain improvements through distillation. This is seen by the regions above the black line.

\subsection{Case Study-3: Ising Simulation}
\begin{figure}[ht]
    \centering
    \includegraphics[width=0.75\columnwidth]{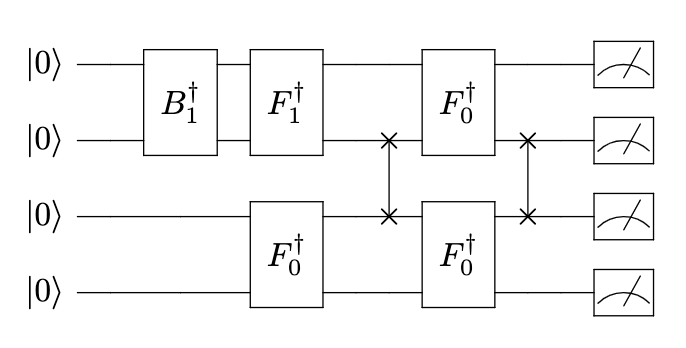} 
    \caption{Quantum circuit that transforms computational basis states into transverse Ising eigenstates. The two qubit gates $F_{1}^{\dagger}$ and $F_{0}^{\dagger}$ apply the inverse Fourier transform and the $B_{1}^{\dagger}$ with the inverse Bogoliubov transformation. Gates represented with crosses correspond with the fermionic SWAP (fSWAP) gates that take care of the fermion anti-commutation relations.}
    \label{fig:ising-circuit}
\end{figure}

\begin{figure}[!t]
    \centering
    \includegraphics[width=0.96\columnwidth]{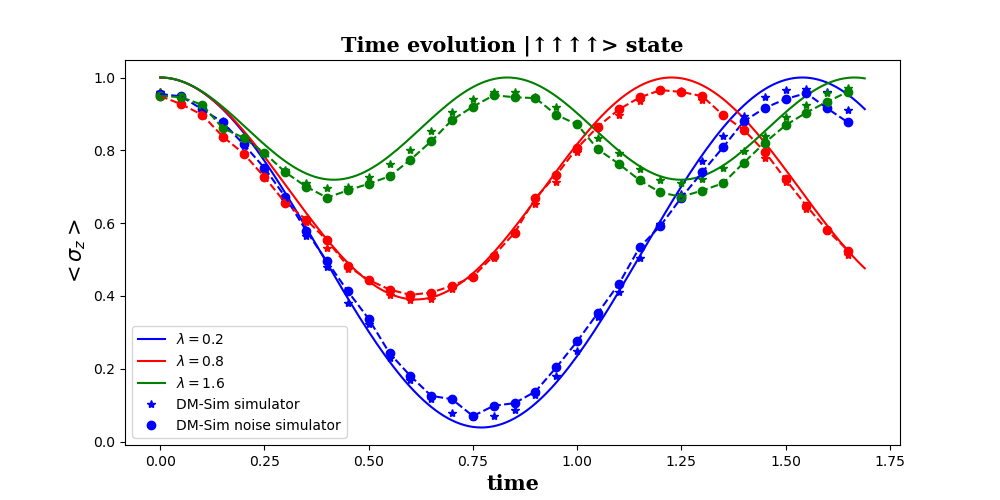} 
    \caption{Time evolution simulation of magnetization $\langle\sigma_z\rangle$ for the state $|\uparrow\uparrow\uparrow\uparrow \rangle$ of a $n=4$ spin state. Solid line represents the exact solution with no-noise in comparison with the experimental simulations on TANQ-Sim with the noise model (scatter points).}
    \label{fig:ising-dmsim}
\end{figure}

The transverse field Ising model (TFIM) is a key simulation problem for quantum computing \cite{PhysRevA.95.052339}. Here we use TANQ-Sim to simulate the time evolution of the TFIM system. Similar to its classical counterpart, we study a linear chain of two-state systems, such as spin-$\frac{1}{2}$  particles, with a specified external field ($\lambda$) and local couplings ($J$) with the following Hamiltonian parameterized as:
\begin{equation}
\mathcal{H}=J \sum_{i=1}^{n}\sigma_{i}^{x}\sigma_{i+1}^{x} + \sigma_{1}^{y}\sigma_{2}^{z}\cdots\sigma_{n-1}^{z}\sigma_{n}^{y} +\lambda\sum_{i=1}^{n}\sigma_{i}^{z},
\end{equation}
This Hamiltonian described in the spin-basis will be diagonalized using standard transformation techniques \cite{LIEB1961407}. For simple experiments, we consider the $n=4$ spin chain and the quantum circuit is shown in Figure~\ref{fig:ising-circuit} revised from \cite{PhysRevA.79.032316, CerveraLierta2018exactisingmodel}.
The time evolution of a quantum system driven by a time-dependent Hamiltonian is described using the Schrodinger equation $|\Psi (t) \rangle  = U(t) |\Psi_0 \rangle$, with the time evolution operator
$U(t) = e^{-i \mathcal{H} t}$,
where $|\Psi_0 \rangle$ is the initial state. For this experiment, we take all spins aligned in the positive $z$ direction as initial state, 
i.e. $|\uparrow \uparrow \uparrow \uparrow \rangle$, which is the $|0000\rangle$ on the computational basis. 
Finally, we can obtain the expectation value of transverse magnetization, i.e., $M_z = \frac{1}{2} \langle \sigma_z \rangle$.
Figure~\ref{fig:ising-dmsim} shows the time evolution of transverse magnetization $\langle \sigma_z \rangle$ with and without noise on TANQ-Sim.

\section{Related Work}

There are multiple ways to represent qubits and simulate quantum circuits classically, including \emph{state-vector}~\cite{jones2019quest, li2021sv}, \emph{density-matrix}~\cite{li2020density, chen2021low}, \emph{tensor-network}~\cite{markov2008simulating, nguyen2021tensor}, \emph{decision diagrams}~\cite{miller2006decision, grurl2020considering}, \emph{stabilizer}~\cite{aaronson2004improved, bravyi2019simulation}, and \emph{device-level simulation} such as \emph{pulse-based} simulation~\cite{mckay2018qiskit}.

We are particularly interested in multi-GPU or multi-accelerator based simulation. Some early works including Zhang et al.~\cite{zhang2015quantum} which presents a state-vector simulator for a single GPU-node incorporating 4 NVIDIA Kepler GPUs. Inter-gate reuse is exploited to mitigate communication overhead. The inter-GPU communication is through CPU host memory where the bandwidth of PCIe severely constraint the performance.

UniQ~\cite{zhang2022uniq} proposes QTensor, a unified intermediate representation for different quantum circuit simulations. For density matrix simulation, different shapes of Tensors are used in the Kraus and gate operations. It also provides a unified hierarchical hardware abstraction to map the simulation workload to various devices. In particular, it leverages the optimization techniques performed in HyQuas~\cite{zhang2021hyquas}, namely TransMM method, OShareMem method, and global-local swap method, to automatically partition a given circuit into different groups and decide the allocation of the workloads on backend GPUs for better performance. While both simulation frameworks (i.e. UniQ and HyQuas) show significant performance benefits, the evaluation is limited to small scale, e.g., single node with 4 V100 GPUs fully connected by NVLink 2.0, and one A100 node that has a single A100 GPU. In contrast, our work aims to design general-purpose, noise-enabled density matrix quantum for large-scale GPU cluster demonstrated on the Perlmutter system with 2,048 A100 GPUs.

DM-Sim~\cite{li2020density} presents a multi-GPU density matrix simulation approach relying on CPU-side all-to-all communication through MPI and GPUDirect RDMA. It also optimizes through single kernel execution to avoid frequent CPU-GPU data movement and context switch. SV-Sim~\cite{li2021sv} presents a multi-GPU state-vector simulator based on the SHMEM communication interface for single kernel execution and fine-grained computation-communication overlapping. It also optimizes through gate-specialized implementation and a functional pointer mechanism. As a comparison, TANQ-Sim also adopts single kernel execution through NVSHMEM for inter-node communication. However, we found that fine-grained communication could lead to considerable overhead with large number of qubits and complex circuits. In TANQ-Sim, we fuse the fine-grained communication as one transaction for a single thread, significantly reducing the overhead. Instead of relying on gate specialization, we propose universal gate approach (i.e., \texttt{C1} and \texttt{C2}) and advanced gate fusion technique to reduce the gates. More importantly, we propose a novel approach to leverage the latest double-precision tensorcores of Ampere and Hopper GPUs for density matrix quantum simulation.

Another recent simulation work is qTask~\cite{huang2023qtask}, which introduces a novel programming model to allow the simulator to incrementally and partially update affected regions of the quantum state without exhaustive simulation. Once the circuit is written with the programming model, qTask divides a state vector into partitions as collective blocks. Each partition spawns one or multiple tasks to perform gate operations on designated memory regions. qTask keeps a list of partitions called frontiers and handles the newly added gates and removed gates accordingly. However, qTask was only evaluated in small scale and was not accelerated by GPUs.

\section{Conclusion}

In this paper, we present TANQ-Sim, a full-scale density matrix based simulation framework for simulating practical deep quantum circuits with both coherent and non-coherent noise. TANQ-Sim is integrated with the Microsoft QIR infrastructure and can support various front-ends through QCOR. It is designed to be accelerated by the Ampere and Hopper GPU double-precision tensorcores, optimized through gate-fusion, and scaled through GPU-side communication using NVSHMEM. Extensive evaluations and case studies on the NERSC Perlmutter supercomputer demonstrate the functionality, performance, and scalability of TANQ-Sim.

\section*{Acknowledgement}

This material is based upon work supported by the U.S. Department of Energy, Office of Science, National Quantum Information Science Research Centers, Quantum Science Center. This research used resources of the National Energy Research Scientific Computing Center (NERSC), a U.S. Department of Energy Office of Science User Facility located at Lawrence Berkeley National Laboratory, operated under Contract No. DE-AC02-05CH11231. This research used resources of the Oak Ridge Leadership Computing Facility, which is a DOE Office of Science User Facility supported under Contract DE-AC05-00OR22725. The Pacific Northwest National Laboratory is operated by Battelle for the U.S. Department of Energy under Contract DE-AC05-76RL01830.

\bibliographystyle{unsrt}
\bibliography{SC23}

\end{document}